# Modelling galaxies with $f(E, L_z)$; a black hole in M32

Walter Dehnen
*Theoretical Physics, 1 Keble Road, Oxford OX1 3NP, United Kingdom*

**ABSTRACT**
A technique for the construction of axisymmetric distribution functions for individual galaxies is presented. It starts from the observed surface brightness distribution, which is deprojected to gain the axisymmetric luminosity density, from which follows the stars' gravitational potential. After adding dark mass components, such as a central black hole, the two-integral distribution function (2I-DF) $f(E, L_z)$, which depends only on the classical integrals of motion in an axisymmetric potential, is constructed using the Richardson-Lucy algorithm. This algorithm proved to be very efficient in finding $f(E, L_z)$ provided the integral equation to be solved has been properly modified. Once the 2I-DF is constructed, its kinematics can be computed and compared with those observed. Many discrepancies may be remedied by altering the assumed inclination angle, mass-to-light ratio, dark components, and odd part of the 2I-DF. Remaining discrepancies may indicate, that the distribution function depends on the non-classical third integral, or is non-axisymmetric.

The method has been applied to the nearby elliptical galaxy M32. A 2I-DF with $\sim 55°$ inclination and a central black hole (or other compact dark mass inside $\sim 1$pc) of $1.6\text{-}2 \times 10^6 M_\odot$ fits the high-spatial-resolution kinematic data of van der Marel et al. remarkably well. 2I-DFs with a significantly less or more massive central dark mass or with edge-on inclination can be ruled out for M32. Predictions are made for observations with the *HST*: spectroscopy using its smallest square aperture of $0.''09 \times 0.''09$ should yield a non-gaussian central velocity profile with broad wings, true and gaussian-fit velocity dispersion of 150–170 kms$^{-1}$ and 120–130 kms$^{-1}$, respectively.

## 1 INTRODUCTION

Since the 1970s it is known that, in general, elliptical galaxies are not simply isotropic oblate stellar systems flattened by rotation, but have triaxial shapes, which are supported by anisotropic velocity distributions. To better understand the dynamical structure of these objects, detailed dynamical models are required, and a natural way of constructing them consists of the following steps. (i) Find a three-dimensional luminosity density $\rho$ which gives the observed surface brightness when projected and convolved with the atmospheric seeing; (ii) assume a mass-to-light ratio and integrate the Poisson equation to obtain the potential of the stars, to which further mass components may be added, e.g. a central black hole or a dark halo; (iii) solve the integral equation $\rho = \int f d^3v$ for the distribution function $f$, which must depend only on the isolating integrals of motion in the given potential to satisfy the collisionless Boltzmann equation; and (iv) finally, compare the kinematic observables with those observed, to test whether the model is consistent or not.

In general, this procedure is very complicated and allows a huge amount of freedom. Already the deprojection of a triaxial body is highly non-unique. However, if one assumes axisymmetry and fixes the inclination, there is de facto only one smooth and physical density distribution for any allowed inclination (Palmer 1994), but strictly speaking the deprojection is unique only for edge-on inclination (Rybicki 1987). Another reason for the assumption of an axisymmetric configuration is its simple phase-space structure easing the dynamical modelling: there is only one major orbit family and two classical integrals of motion, while triaxial potentials allow for four major orbit families with, in general, the energy being the only classical invariant. Additionally, there is also some observational evidence, that elliptical galaxies in their majority have near-oblate shapes, as indicated by the alignment of kinematic and photometric axes in many of these objects (Franx, Illingworth & de Zeeuw 1991). Another indication comes from studies of gas flows, for instance, in the case of the elliptical galaxy IC 2006 the form and velocity field of an equatorial gas ring constrains the potential to be round in the equatorial plane (Franx, van Gorkom & de Zeeuw 1994). This view is supported by recent results of numerical $N$-body simulations with small amounts of gas being included: they predict near-oblate shapes both for remnants of disk mergers (Barnes & Hernquist 1994) and for dark matter halos in CDM cosmogonies (e.g. Katz & Gunn 1991).

Typical stellar orbits in axisymmetric potentials obey three integrals of motion, namely the energy $E$, the $z$-component of angular momentum $L_z$, and the non-classical, third integral $I_3$. Unfortunately, the third integral is generally not known explicitly and must be evaluated by perturbation theory or similar means. Therefore, one often restricts oneself to models with stellar distribution functions (DFs) depending only on the two classical integrals of motion, i.e. $f = f(E, L_z)$ (2I-DF). Though there is no obvious physical reason for the DFs of axisymmetric galaxies to ignore the third integral, such models are useful for the following reasons. (i) The part of the DF even in $L_z$, $f_e$, follows uniquely from density and potential, and hence from the observed surface brightness and the assumptions made on inclination, mass-to-light ratio, and dark mass components. This substantially reduces the amount of freedom in the DF. (ii) Since a two-integral model (2I-model) is simply to construct (compared to any 3I-model), it can be used as a template, i.e. the comparison of a galaxy's kinematics with that predicted by the 2I-model already gives clear hints as to the true nature of the dynamical configuration of that galaxy. Instead of modelling the DF, one may also consider its velocity moments, i.e. streaming velocity and velocity dispersion, which in the case of $f = f(E, L_z)$ follow uniquely from the Jeans equations. However, there is always the possibility that the underlying DF is not positive definite, which may be undetectable from its moments. Moreover, nowadays not only velocity and dispersion of the stellar line-of-sight velocity profiles (VPs), but also their shapes can be extracted from galaxy spectra (Bender 1990; Rix & White 1992; van der Marel & Franx 1993; Winsall & Freeman 1993; Kuijken & Merrifield 1993). To fully exploit these data one must either calculate the DF or solve the Jeans equations for the higher order ($\sim 16$) velocity moments (Magorrian & Binney 1994).

In this paper I present a numerical method for the construction of two-integral distribution functions for *individual* galaxies. These allow to explore the significance of specific observational features for the dynamical structure of a galaxy. Thus such models grant deeper insights than do simple spheroidal models as those of Dehnen & Gerhard (1994) or Qian et al. (1994). The method essentially consists of (1) a deprojection of the surface brightness into an axisymmetric luminosity density and (2) the construction of the corresponding 2I-DF with the assumption of some mass-to-light ratio, which may include dark mass components such as a massive central black hole or a dark halo. Both steps are similar insofar as a linear integral equation has to be solved. In the given method this is done using the Richardson (1972) –Lucy (1974) (RL) algorithm, which – after some modifications – turns out to be very efficient in solving both equations.

The method is illustrated by an application to the nearby elliptical M32, which is suspected harbouring a massive central black hole (Tonry 1987; Dressler & Richstone 1988; Richstone, Bower & Dressler 1990). Recently, Qian et al. (1994) have employed the method of Hunter & Qian (1993) to compute $f(E, L_z)$ for a simple mass model for the central $\sim 20''$ of M32 used by van der Marel et al. (1994b) in modelling the velocity moments. Although the model presented here differs from the Qian et al. analysis in details of the stellar mass distribution and in the technique for the evaluation of $f(E, L_z)$, the main result is the same: a 2I-model with a 1.6-2 $\times 10^6 M_\odot$ black hole gives a good fit to the kinematic data observed by van der Marel et al. (1994a). However, some small discrepancies between the VPs of M32 and of the models remain, which suggest that the DF may depend weakly on the third integral.

The sections of this paper are designed to be read almost independently from each other. Their contents are as follows. Section 2 describes how the RL algorithm can be used to find $f(E, L_z)$ from density and potential; numerical details and the result of a test with an analytical model are given. The construction of two-integral models for individual galaxies and the comparison of their kinematics with that observed is outlined in Section 3. As an application Section 4 deals with 2I-models for M32, and Section 5 sums up and concludes. Appendix A briefly describes the RL algorithm and how to achieve faster convergence; Appendix B contains formulae for the recovery of $f(E, L_z)$ from the derivatives of the density; and Appendix C gives a Monte-Carlo procedure for the evaluation of VPs accounting for binning and seeing.

## 2 NUMERICALLY RECOVERING $f(E, L_z)$ FROM DENSITY AND POTENTIAL

For an axisymmetric stellar system with a 2I-DF the spatial density $\rho$ is related to the phase space density $f$ by

$$\rho(\Psi, R) = 2\pi \int_0^\Psi dE \int_0^{2R^2(\Psi - E)} \frac{dL_z^2}{L_z R} f_e(E, L_z^2), \quad (1)$$

where the density is written as a function of the cylindrical radius $R$ and the relative potential $\Psi$, while $f_e$ denotes the part of the 2I-DF even in $L_z$, which only contributes to the density. It has been known for more than 30 years that the solution of this equation for $f_e(E, L_z^2)$ given $\rho(\Psi, R)$ is unique (Lynden-Bell 1962), however, only recently has a general inversion formula been found (Hunter & Qian 1993), that is the generalization of Eddingtons (1916) formula for the $f(E)$ generating a spherical system. This inversion formula requires the knowledge of $\partial \rho / \partial \Psi$ in parts of the complex $\Psi$-plane, which can be numerically evaluated for many functional forms of $\rho$, even if it is not expressed by elementary functions in $\Psi$ and $R$ (Qian et al. 1994). However, this technique can hardly be applied to densities in tabulated form, since (i) the evaluation of $\partial \rho / \partial \Psi$ is almost impossible with high accuracy, and (ii) the density may well be contaminated with noise, which is amplified to strong oscillations in the DF.

In this section a scheme to solve for $f_e(E, L_z^2)$ will be given that is designed for tabulated density fields, guarantees a physical DF, and nevertheless is fast and accurate.

## 2.1 Using the Richardson-Lucy algorithm

A suitable tool for the recovery of $f_e(E, L_z^2)$ for general axisymmetric density fields is the Richardson (1972) –Lucy (1974) algorithm (hereafter RL algorithm). It was first introduced into stellar dynamics by Newton & Binney (1987), who used it to construct a positive spherical DF for M87; later on Gerhard (1991) applied it for the evaluation of anisotropic DFs in spherical potentials. A description of this algorithm, and how to accelerate and modify it for faster convergence, is given in Appendix A. Starting from any guess for $f_e(E, L_z^2)$, one first projects the DF into $\rho(\Psi, R)$ according to equation (1), and then corrects $f_e(E, L_z^2)$ using the discrepancy between the actual density and that of the DF. This pair of steps may then be repeated until satisfactory convergence is obtained. Applying the RL algorithm after multiplying equation (1) with $R^{-s}\rho^{-p}$ on both sides (Appendix A2) yields for the correction step

$$\frac{f_{e,n+1}(E, L_z^2)}{f_{e,n}(E, L_z^2)} = \frac{\int_{\Psi_1}^{\Psi_2} d\Psi \int_{\frac{L_z}{\sqrt{2(\Psi-E)}}}^{R_\Psi} \frac{dR}{R^{s+1}} \frac{\rho(\Psi, R)^{1-p}}{\rho_n(\Psi, R)}}{\int_{\Psi_1}^{\Psi_2} d\Psi \int_{\frac{L_z}{\sqrt{2(\Psi-E)}}}^{R_\Psi} \frac{dR}{R^{s+1}} \rho(\Psi, R)^{-p}}. \quad (2)$$

The numbers $s$ and $p$ may be chosen to optimize convergence; $R_\Psi$ is given by $\Psi(R_\Psi, 0) = \Psi$; $\Psi_1$ and $\Psi_2$ are the roots of $\Psi = E + L_z^2/(2R_\Psi^2)$; and $\rho_n(\Psi, R)$ is the density generated by $f_{e,n}(E, L_z^2)$, the DF after the $n$-th iteration step. For $L_z = L_{\text{circ}}(E)$ the double integral shrinks to that point to which the circular orbit contributes. For $L_z = 0$ the integral over $R$ diverges at $R = 0$. However, one may apply the RL algorithm to $\rho(\Psi, 0) = 4\pi \int_0^\Psi dE \sqrt{2(\Psi-E)} f(E, 0)$ to get

$$\frac{f_{e,n+1}(E, 0)}{f_{e,n}(E, 0)} = \frac{\int_E^{\Psi(0,0)} d\Psi \sqrt{\Psi-E} \, \rho(\Psi, 0)^{1-p}/\rho_n(\Psi, 0)}{\int_E^{\Psi(0,0)} d\Psi \sqrt{\Psi-E} \, \rho(\Psi, 0)^{-p}}. \quad (3)$$

Thus, recovering $f_e(E, L_z^2)$ from $\rho(\Psi, R)$ via the RL algorithm yields a procedure, which at each iterative step requires a two-dimensional integration on almost each point of the two-dimensional grids in $(E, L_z^2)$ and $(\Psi, R)$. Appendix B describes procedures, which require only one-dimensional integrations at each point of the two-dimensional grids. However these involve taking the derivative with respect to $\Psi$ or $R$ on both sides of equation (1), while my goal is to recover $f_e(E, L_z^2)$ from a tabulated density, for which these derivatives cannot be evaluated with sufficient accuracy.

## 2.2 Unphysical distribution functions

There may well be combinations of positive density and potential for which no underlying non-negative $f_e(E, L_z^2)$ exists. Then the scheme given above will never converge to $\rho \equiv \rho_n$ (within the numerical accuracy) as $n$ increases. The remaining discrepancies in the density may either by confined to spatial scales as small as the resolution element of the grid used to represent the density, or they occur on a larger scale. In the first case, the discrepancies may easily be interpreted as noise, since the RL algorithm guarantees a positive DF, and hence gives a smooth density.

In the second case, the large scale behaviour of density and potential obliges the true $f_e(E, L_z^2)$ to be somewhere negative. When using the RL scheme, this will be recognized by the fact, that the error in the density does not converge to zero, despite the DF continuously decreases in a certain region of phase space. In such a case, the above scheme can be used to recover the true but non-physical $f_e(E, L_z^2)$ by applying it a second time on the density $\tilde{\rho} = k\rho_n - \rho$, where $k$ has to be chosen to keep $\tilde{\rho}$ everywhere positive, and finally constructing the non-physical DF as difference of two physical ones.

## 2.3 Numerical implementation

For the numerical implementation of the procedure, $\rho(\Psi, R)$ is tabulated on a grid logarithmic in $R_\Psi$ and equidistant in $R/R_\Psi$, whereas $f_e(E, L_z^2)$ is tabulated on a grid logarithmic in $R_E$, where $\Psi(R_E, 0) = E$, and equidistant in $L_z^2/L_{\text{circ}}^2(E)$. Interpolations are done in $\ln\rho$ and $\ln f$ using cubic splines. The integrals $dL_z^2$ and $dR$ in equations (1) and (2) are computed by integrating the spline curves exactly, while those over $E$ and $\Psi$ are evaluated by gaussian quadratures.

Starting from a density field $\rho(R, z)$ and a potential $\Psi(R, z)$, the routine first eliminates $z$ to yield $\rho(\Psi, R)$, computes $\Psi_1$ and $\Psi_2$ for the above grid in $(E, L_z^2)$, and pre-calculates the integrals in the denominators of equations (2) and (3) on that grid. An initial $f(E, 0)$ is guessed using Eddingtons (1916) formula for isotropic spherical systems, which is also valid for $f(E, 0)$ of axisymmetric systems. The initial guess for $f_e(E, L_z^2)$ is then obtained by first using the RL algorithm to fully recover $f(E, 0)$ (equation 3) and then multiplying it with some function of $L_z^2/L_{\text{circ}}^2$. The convergence of the iteration is accelerated as described in Appendix A3. After each correction step, $f_{e,n}$ is smoothed on the scale of the grid cells, to avoid fitting small-scale noise. With a grid of 80×11 points for the density, 131×17 points for the DF, and 60 points for the gaussian quadratures, each iteration takes about 40 seconds CPU time on a DEC Alpha 3000-600 workstation.

## 2.4 Testing with a simple model

Recently, Evans (1994) found a family of axisymmetric models with simple analytical 2I-DFs. In these models the potential is proportional to $(R_c^2 + R^2 + z^2 q^{-2})^{-\beta/2}$, and the 2I-DFs are always finite sums of powers in $E$ and $L_z^2$. To test the algorithm and its numerical implementation, I chose the model with $\beta = 0.5$ and $q = 0.93$. The numerical apparatus was fed by the density field tabulated between $0.005 R_c$ and $200 R_c$, the potential was obtained by multipole expansion. After only three iterations the routine found $f_e(E, L_z^2)$, such that the RMS relative deviation between its density and the input was as small as 0.003. Figure 1 compares the exact DF (solid contours) with the numerical results (dashed), the differences are almost invisible.

## 3 MODELLING INDIVIDUAL GALAXIES WITH TWO-INTEGRAL MODELS

This section describes how two-integral models for individual galaxies can be obtained and compared to the observations. A first application follows in Section 4 with the E3 galaxy M32.

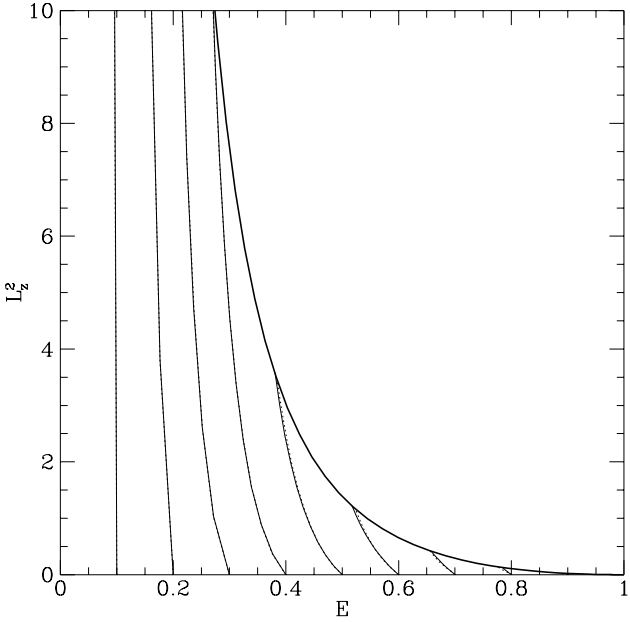

**Figure 1.** Contour lines of $f_e(E, L_z^2)$ for Evans' (1994) power-law model with $\beta = 0.5$ and $q = 0.93$; solid: analytical DF, dotted: DF numerically reconstructed via the RL algorithm.

### 3.1 The mass model

There are two fundamentally distinct ways to derive a three-dimensional luminosity density from a given surface brightness: fitting some parameterized function and using unparameterized methods. Recently, Merritt & Tremblay (1994) have discussed the relative merits of both techniques. In the present framework, a fitted parameterized function would be free of small scale noise and advantageous in computing the gravitational potential, however, the density would always be biased. To allow for general forms of $\rho(R, z)$ (i.e. any radial run, disky/boxy contours, changing ellipticity), I use an (almost) unbiased technique: the RL algorithm, which was introduced for this purpose by Binney, Davies & Illingworth (1990). Differently from these authors, I modify the correction equation with $\alpha = I^{-p}$ (in the notation of Appendix A2), to yield a better convergence. Then the projection and correction equation are

$$I_n(x, y) = \frac{1}{\sin i} \int_{-\infty}^{\infty} ds\, \rho_n(\sqrt{s^2 + x^2},\, y \sec i + s \cot i) \quad (4)$$

$$\frac{\rho_{n+1}(R, z)}{\rho_n(R, z)} = \frac{\int_{-\pi/2}^{\pi/2} dt\, \tilde{I}_n(R\cos t,\, z\sin i + R\cos i \sin t)}{\int_{-\pi/2}^{\pi/2} dt\, [I(R\cos t,\, z\sin i + R\cos i \sin t)]^{-p}} \quad (5)$$

with $\tilde{I}_n(x, y) \equiv I(x, y)^{1-p}/I_n(x, y)$, where $x$ and $y$ are respectively the co-ordinates in the plane of sky along the major and minor axes of the galaxy. The angle of inclination is denoted $i$ with $90°$ meaning edge-on projection. The choice $p = 0$, for which the above equations reduce to those already given by Binney et al., turned out to give best convergence in the case of a centrally flat surface brightness, while for a cusp $p \approx 1$ is more efficient. As initial guess for the density I use the spheroidal model:

$$\rho(R, z) = \rho_0 m^{-\gamma} (m + 1)^{\gamma - \beta}; \quad m^2 = (R^2 + z^2 q^{-2})/r_0^2, \quad (6)$$

which has surface brightness

$$I(x, y) = \frac{4 q r_0 \rho_0}{q_p (\xi + 1)^{\beta - \gamma - 1/2}} \int_0^1 \frac{(\xi + t^2)^{1-\gamma}(1 - t^2)^{\beta - 2}\, dt}{\sqrt{2\xi + (1 - \xi)t^2}} \quad (7)$$

$$\xi^2 = (x^2 + y^2 q_p^{-2})/r_0^2, \qquad q_p^2 = q^2 \sin^2 i + \cos^2 i.$$

The parameters $\gamma$, $\beta$, $\rho_0$ and $r_0$ are fitted to the surface density along the major axis, the axis ratio $q$ is guessed from a mean value for the apparent axis ratio $q_p$ and the inclination.

In order to ensure, that the errors are not systematically distributed, i.e. that the result is not biased towards the starting density, a minimum number of iterations is necessary. To avoid an amplification of noise in the data, the density field may be smoothed after each correction step on a scale similar to the resolution element of the grid.

Finally the stellar mass distribution is obtained by multiplying the luminosity density with a (usually constant) mass-to-light ratio $\Upsilon$, and the stars gravitational potential can be evaluated, e.g. by using multipole expansion (cf. Binney & Tremaine 1987, §2.4). Additional dark mass components may be added to the potential.

### 3.2 The velocity profiles

After the construction of $f_e$, as outlined in Section 2, one finally can compute its observables. The complete set of kinematic observables for a stellar system are the *velocity profiles* (hereafter VP) at each position on the sky. The VP $l(v_{\rm los}; x, y)$ is simply the phase space distribution function integrated over the non-observable co-ordinates, which are the stars' positions $s$ along the line of sight and their velocities in the plane of the sky

$$l(v_{\rm los}; x, y) = \int_{-\infty}^{\infty} ds \int dv_x \int dv_y f(E, L_z). \quad (8)$$

Robust algorithms to extract the VPs from observed galaxy spectra have been available for a few years (Bender 1990; Rix & White 1992; van der Marel & Franx 1993; Kuijken & Merrifield 1993; Winsall & Freeman 1993). In general, spectra are only measured at a few slit positions, i.e. major and minor axes, and only in the central regions of galaxies are the signal-to-noise ratios high enough to extract the VPs. In order to compare the model with such observations, one must account for effects of the finite slit width and pixel size as well as for atmospheric seeing. All these will lead to some averaging and can change the VPs near the center of a galaxy dramatically. Appendix C gives a numerical method for the computation of VPs of 2I-models based on Monte Carlo methods, which allow these effects to be included.

### 3.3 Comparing with the observed kinematics

Because of the uniqueness of $f_e$, it is useful, to first compare its observables with those of the modelled galaxy and if necessary correct the inclination or mass-to-light ratio before calculating $f_o$, the DF's odd part. The observables of $f_e$ are simply the even parts of the VPs, $VP_e$, and its moments[1] especially the RMS-velocity $\langle v_{los}^2 \rangle^{1/2}$.

The RMS-velocity scales with the square root of the (constant) mass-to-light ratio $\Upsilon$, while the inclination influences the ratio of RMS-velocities on the minor and major axes $\langle v_{los}^2 \rangle_{min}^{1/2} / \langle v_{los}^2 \rangle_{maj}^{1/2}$. The latter decreases with decreasing $i < 90°$ (Binney et al. 1990; Dehnen & Gerhard 1994). A central black hole alters the kinematics in the central regions by causing $\langle v_{los}^2 \rangle^{1/2}$ to rise as $r^{-1/2}$ (neglecting seeing), while a dark halo alters the radial run of $\langle v_{los}^2 \rangle^{1/2}$ at large radii. The behaviour with inclination implies, that a galaxy with a greater ratio $\langle v_{los}^2 \rangle_{min}^{1/2} / \langle v_{los}^2 \rangle_{maj}^{1/2}$ than a 2I-model with assumed edge-on projection cannot have a 2I-DF, and therefore must be described by a three-integral model[2].

After inclination, mass-to-light ratio and any dark mass components have been adapted to give $\langle v_{los}^2 \rangle^{1/2}$ in agreement with the observations, the shapes of the $VP_e$ can be compared with the observed ones, which again may rule out a 2I-DF. Once $f_e$ has been determined, the remaining job is to test, whether there exists a $f_o$ that (i) leads to a physical DF and (ii) results in VPs in agreement with the observations. One constraint is that the observed mean velocity $\langle v_{los} \rangle$ is nowhere greater than that of the complete rotating model in which all stars rotate in the same sense.

## 4 MODELLING M32

Because of its high central density and steep central gradients in rotation and velocity dispersion, the nearby compact elliptical M32 is believed to lodge a massive black hole (Tonry 1987; Dressler & Richstone 1988; Richstone, Bower & Dressler 1990). However, this presumption was based on spherical mass models, despite M32 being as flat as E3. Recently, high spatial resolution data have been obtained both for the surface brightness by *HST* observations (Lauer et al. 1992) and for the kinematics (VPs) by ground based spectroscopy (van der Marel et al. 1994a, hereafter vdM94a). These data make it worthwhile to investigate the dynamics of M32 with self-consistent distribution-function models, as described above. Very recently, van der Marel et al. (1994b) have used a simple stellar mass model fitting the *HST* data to solve the Jeans equations up to third order assuming a 2I-DF, while Qian et al. (1994) computed the actual DF for that model using the Hunter & Qian (1993) technique. I discuss the relation to this work in 4.6.

### 4.1 The luminosity density

Figure 2 shows the surface brightness and ellipticity profiles from which an axisymmetric luminosity density was derived. Fig. 2 combines the *HST* observations of Lauer et al. (1992) for the inner $4''$ (using $V$-$R = 0.4$ mag), the data of Peletier (1993) out to $32''$, and Kent's (1987) data further out. Inside

---

[1] This also holds for axisymmetric three-integral models provided the third integral is even in $v_\phi$.
[2] An exception is the case in which the dominating mass component is flatter than the stellar distribution, which reduces the need of near-circular orbits to flatten the stellar system, and hence in projection results in less motion on the major axis.

**Table 1.** Deprojection of the surface brightness of M32

| $i$ | E | $\gamma$ | $\beta$ | $N_{RL}$ | $\chi_I$ [mag] |
|---|---|---|---|---|---|
| 90° | E2.7 | 1.53 | 3.3 | 4 | 0.0057 |
| 55° | E4.5 | 1.53 | 3.3 | 5 | 0.0078 |

The first two columns give inclination and intrinsic ellipticity; $\gamma$ and $\beta$ are the parameters of the first guess for the density (equation 6). $N_{RL}$ is the number of RL iterations, while $\chi_I$ gives the RMS relative deviation in surface brightness.

**Table 2.** The models: mass model and even part of the DF

| model | $i$ | $\Upsilon_R$ | $M_\bullet$ | $N_{RL}$ | $\chi_\rho$ |
|---|---|---|---|---|---|
| A | 90° | 2.4 | – | 18 | 0.0067 |
| B | 55° | 2.6 | – | 18 | 0.0065 |
| C | 55° | 2.6 | $1.63 \times 10^6$ | 21 | 0.0062 |
| D | 55° | 2.6 | $1.95 \times 10^6$ | 21 | 0.0063 |

The first column labels the models; $i$ is the inclination angle, $\Upsilon_R$ and $M_\bullet$ are the stellar mass-to-light ratio in $R$ and the mass of the central black hole, respectively, both in solar units. $N_{RL}$ is the number of RL iterations performed, while $\chi_\rho$ gives the RMS relative deviation in the density.

$0\rlap{.}''15$ a power law cusp $I \propto r^{-0.53}$ with axis ratio 0.73 is assumed, which is consistent with the *HST* data (Lauer et al. 1992). The deprojection has been carried out as explained in 3.1 with the isophotes assumed to be exactly elliptical. Two angles of inclination have been considered: $i = 90°$ (edge-on) and $i = 55°$. The details of the deprojections are given in Table 1. By more iterations the deviations from the input luminosity could be further reduced, however, this does not seem wise since the measured luminosity is hardly more accurate than 0.01 magnitudes.

### 4.2 Constructing $f_e(E, L_z)$

In the case of $i = 55°$ $f_e$ was constructed for three values of $M_\bullet$, the mass of the central black hole: 0, 1.63, and 1.95 million solar masses, while for $i = 90°$ the galaxy was modelled with $M_\bullet = 0$ only (see Table 2). All models have constant stellar mass-to-light ratio. The differences between $\rho$ generated by the DF and that obtained from the deprojection are of the order of (or smaller than) the uncertainty of the density due to incomplete deprojection and erroneous surface brightness (see above).

For models B and C (Table 2) Figure 3 shows $f_e$ as a function of $R_E$ for various fixed values of $L_z^2/L_{circ}^2$. The asymptotic behaviour with energy at small and large radii can easily be understood. In the envelope a power law for the density of $\rho \propto r^{-3.8}$ is adopted, resulting in a scaling of $f \propto R_E^{-2.3}$ in the keplerian regime. In the centre the density scales as $r^{-1.53}$, which in the case of a black hole (keplerian again) gives $f \propto R_E^{-0.03}$, while for a self-consistent cusp one finds $f \propto R_E^{-2.235}$ (Dehnen 1993).

The $L_z$-dependence of the DFs can be understood by locally comparing the flattening of density and potential: the flatter the density compared to the potential and the steeper the density profile, the more high-angular momentum orbits must be occupied (Dehnen & Gerhard 1994; Qian et al. 1994). In the envelope of both models of Fig. 3 the potential is dominated by its monopole and almost round while the density is moderately flattened (axis ratio ≈ 0.81) and steeply falling off. In the centre of model B (no black hole) the potential is flattened and the density is rather flat (axis ratio ≈ 0.55) but has a shallower gradient than in the envelope. All together lead to a weaker $L_z$-dependence in the inner parts of model B than in its outer parts. The opposite is true for model C with a central black hole, since the latter makes $\Psi$ spherical in the inner parsec and makes many high-$L_z$ orbits necessary.

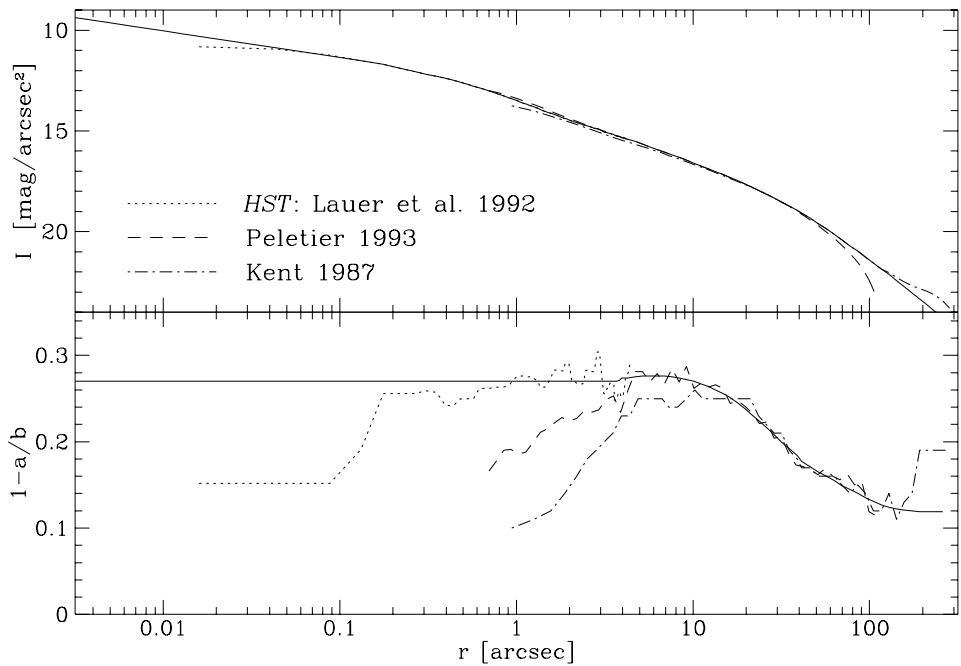

**Figure 2.** Profiles of surface brightness (in the $R$ band) and ellipticity used for modelling M32 (solid), which are combined from various data.

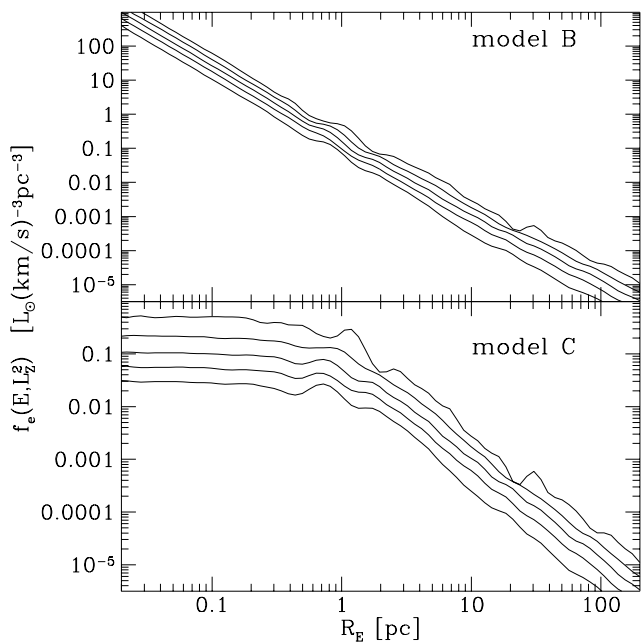

**Figure 3.** Even part of the 2I-DFs for models B and C plotted vs. $R_E$, defined by $\Psi(R_E, 0) = E$, for $L_z^2/L_{\rm circ}^2 = 0$, 0.25, 0.5, 0.75, 1 (from bottom to top).

The DFs resemble each other in their small-scale behaviour, which depends on the small-scale behaviour of $\rho(\Psi, R)$. There are two clear signatures. First, the wiggle at $R_E \approx 1$ pc corresponds to the turn-over of density and surface brightness to a power law inside $0\rlap{.}''15$. This gives a drastic change in $\partial\rho/\partial\Psi$, what is closely related to the DF's run with energy, see Hunter & Qian's (1993) formula for $f(E, L_z)$. A central power law in the density may have evolved during the adiabatic grow of a black hole (Young 1980; Quinlan, Hernquist & Sigurdsson 1994). Therefore, the coincidence of the corresponding pattern in the DF with the energy, at which the DF changes its slope due to the presence of a black hole, may be not accidental (note that $M_\bullet$ is solely fitted to the kinematics, see below). However, in case of a adiabatically grown black hole, the DF should turn smoothly from one regime into the other, implying that the radial turn-over in the mass models used here is somewhat too drastic. Second, the wiggle at $R_E \approx 20$ pc, which only occurs at $L_z \approx L_{\rm circ}$, is related to the sudden increase in ellipticity at $\sim 4''$ (Fig. 2), and hence may be somewhat artificial.

### 4.3 The observables of $f_e(E, L_z)$

VdM94a have measured the VPs for several slit positions and fitted a Gauss-Hermite series to each of them (van der Marel & Franx 1993), from which reliable estimates for the RMS-velocities and the even parts of the velocity profiles, VP$_e$, can be achieved. Fig. 4 compares the RMS-velocities with those of the constructed models, while Fig. 5 compares the VP$_e$ at some selected position on the sky. They have been computed using a Monte-Carlo technique, which takes into account effects of binning and seeing (Appendix C). The (constant) mass-to-light ratios $\Upsilon_R$ of the different models

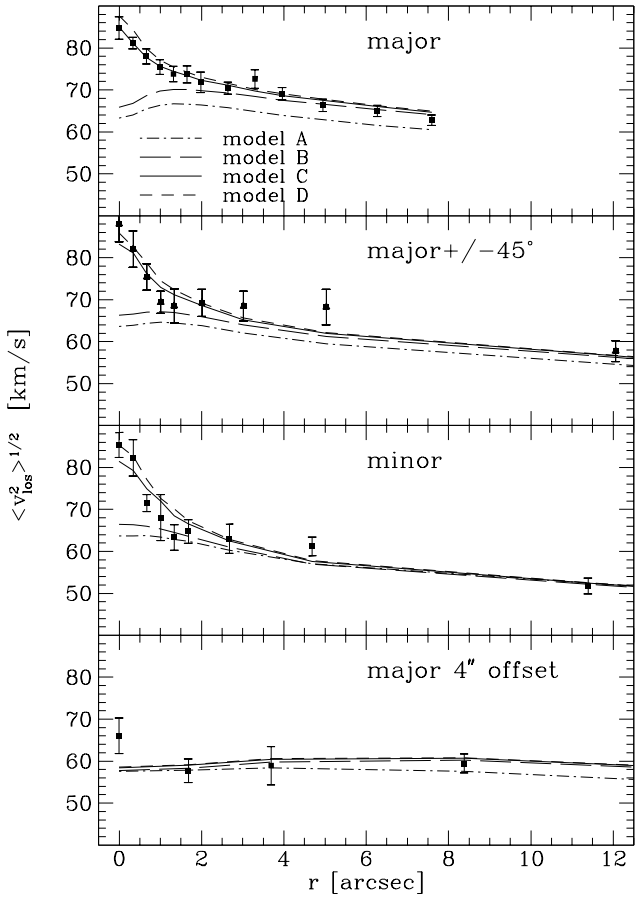

**Figure 4.** RMS-velocities of M32 on four slit positions as observed by vdM94a and of the 2I-models (see Table 2).

have been fixed to match the data on the minor axis. The following points regard further attention:

First, the edge-on model (A) cannot match the kinematics of M32: its $VP_e$ at $5''$ on the major axis is insufficiently flat-topped (Fig. 5), and on the major axis it shows less RMS-motion than M32 (Fig. 4). The latter is untypical for elliptical galaxies, which usually have a somewhat greater ratio $\langle v_{los}^2\rangle_{min}^{1/2}/\langle v_{los}^2\rangle_{maj}^{1/2}$ than an 2I-model seen edge-on (van der Marel 1991). The models with $i = 55°$ show both correctly the ratio of minor to major axis motion and the shape of $VP_e$ outside the centre.

Second, the models without a black hole are unable to reproduce the high velocities inside $1''$ (Fig. 4); they even have centrally declining $\langle v_{los}^2\rangle^{1/2}$, as is typical for self-consistent stellar cusps (Binney 1980; Dehnen 1993). On the other hand, the models with central point masses can account for the measured high central $\langle v_{los}^2\rangle^{1/2}$ (in particular, the central VP is in excellent agreement with model D – see Fig. 5). However, they have slightly more RMS-motion at projected radii of 1-2$''$. This is clearest on the minor axis but not very significant if one compares the $VP_e$ in Fig. 5. A 2I-model with a less massive black hole would have the correct $\langle v_{los}^2\rangle^{1/2}$ at 1-2$''$, but would be unable to account for the high observed velocities in the very centre. Hence, two-integral models cannot reproduce this effect, which therefore is a hint that the DF depends on the third integral.

Third, at about 3-5$''$ on the minor and intermediate axes the models show less RMS-velocity than observed. However, in the $VP_e$ (Fig. 5) this appears to be only a marginal effect, and does not occur on the major axis, where the errors are smallest.

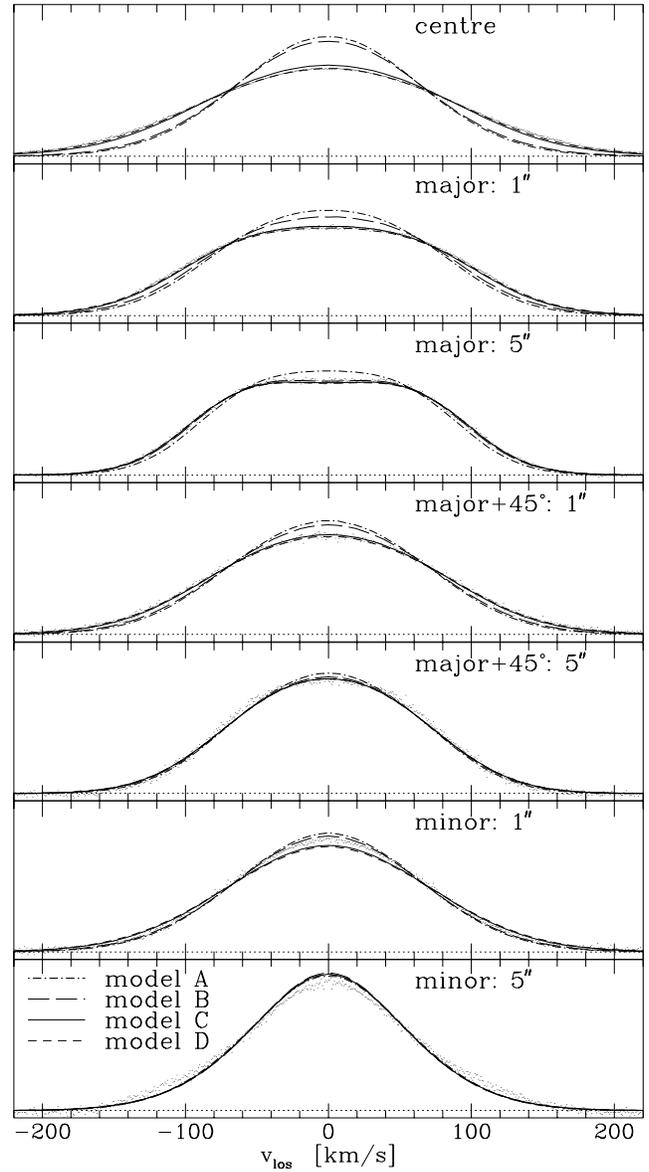

**Figure 5.** Comparing the even parts of the VPs (normalized to unit area) in the centre, and at $1''$ and $5''$ on the major, intermediate and minor axes. The VPs of M32 are represented by points, which have been drawn from the the Gauss-Hermite fit by vdM94a assuming that the given errors are uncorrelated, normally distributed, and are the only errors. In the upper three panels most of these points are overlayed by the VPs of model C or D.

### 4.4 The observables of complete distribution functions

Instead of trying to find the odd part for the DF that results in the best agreement of the predicted and observed VPs, I

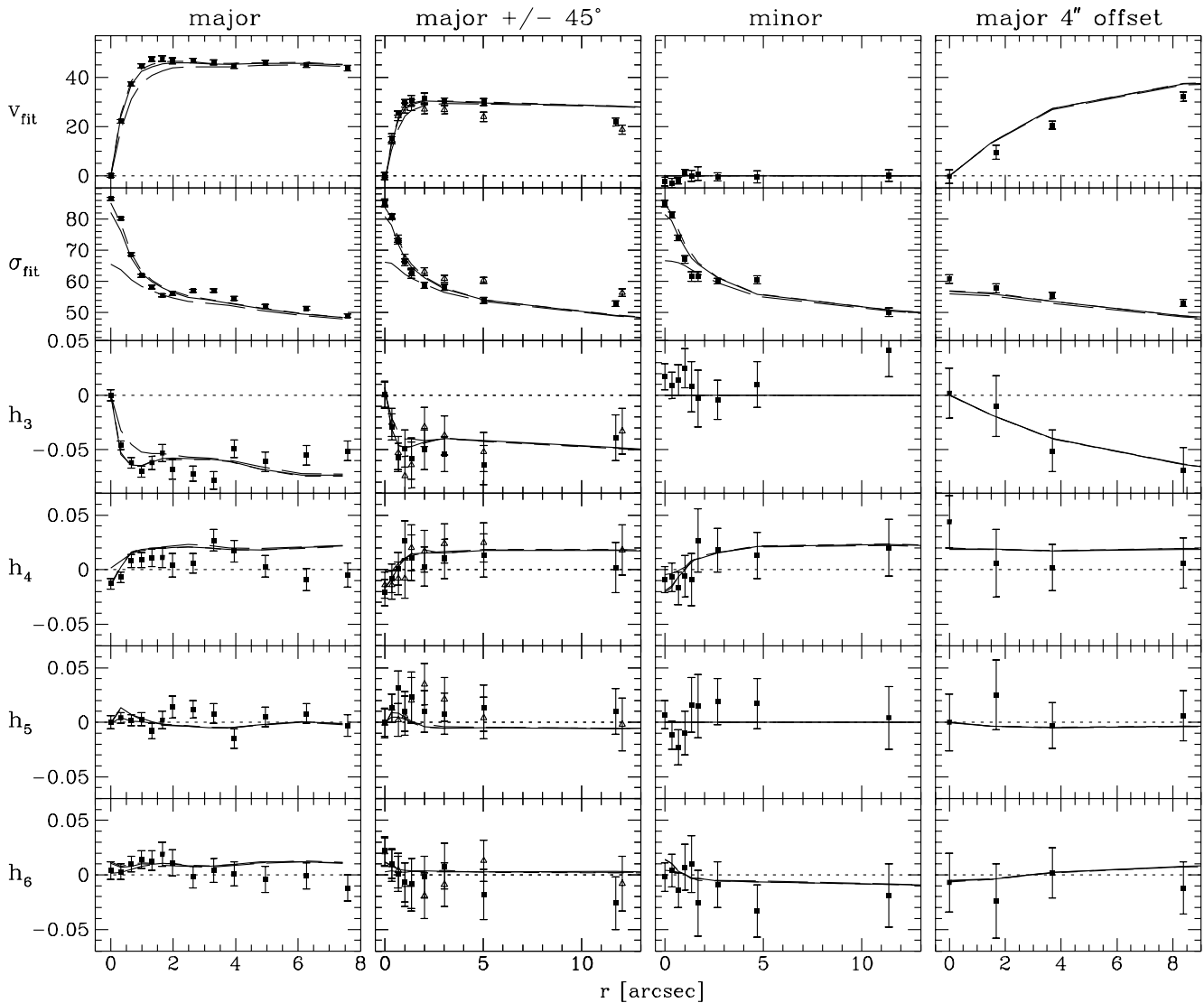

**Figure 6.** Parameters of Gauss-Hermite fits to the VPs of M32 measured by vdM94a. From top to bottom: mean and dispersion of the best-fitting Gaussian to the VP, and the Gauss-Hermite coefficients up to order six. The curves show the corresponding values of the models C (solid) and D (short dashed), which lodge black holes of masses $1.63 \times 10^6 M_\odot$ and $1.95 \times 10^6 M_\odot$, respectively, and of model B (long dashed) without black hole. The atmospheric seeing as well as the spatial binning of the data have been simulated.

make the simple ansatz

$$f(E, L_z) = \begin{cases} e^{\alpha L_z / L_{\rm circ}} f_e(E, 0) & L_z \leqslant 0 \\ 2f_e(E, L_z^2) - e^{\alpha L_z / L_{\rm circ}} f_e(E, 0) & L_z \geqslant 0, \end{cases} \quad (9)$$

$$\alpha = \alpha_1 + (\alpha_2 - \alpha_1) \frac{R_E^2}{R_E^2 + R_0^2}, \quad (10)$$

where $\alpha_1$, $\alpha_2$ and $R_0$ are free parameters. This ansatz implies, that the DF falls off exponentially with $L_z/L_{\rm circ}$ for retrograde orbits with a scale $\alpha^{-1}$ that depends on energy in a very simple form: $\alpha = \alpha_1$ for $R_E \ll R_0$ and $\alpha = \alpha_2$ for $R_E \gg R_0$ with a smooth transition. After a few experiments, I found the models with $\alpha_1 = 6$, $\alpha_1 = 1.7$, and $R_0 = 8\,{\rm pc}$ to give kinematics in reasonable agreement with the data of vdM94a. Figure 6 compares the Gauss-Hermite-fit parameters (van der Marel & Franx 1993) $v_{\rm fit}$, $\sigma_{\rm fit}$ and $h_3, \ldots, h_6$ measured for M32 by vdM94a with those of the models B, C and D, while Figure 7 shows the VPs along the major axis of the models C and D (those with a central black hole) and of M32, as reconstructed from the Gauss-Hermite-fit parameters and their formal errors.

Even for the simple form of $f_{\rm o}$ given by equation (9), the agreement of the models containing a black hole with the VPs of M32 is remarkable, much better than one would expect based on the large variety of three-integral distribution functions and their observables, that are possible in an oblate stellar system (Dehnen & Gerhard 1993). The following points are worth noting.

First, beside leading to the high central rotation and dispersion, the black hole also influences the VP-shapes, which is obvious especially from the $h_i$-profiles on the major axis: the dips in $h_3$ at $1''$ and in $h_4$ in the very centre are not present in the model without central point mass. The negative $h_4$ for the central VP is not what one would naively expect, since the fast motions near the black hole should create broad high-velocity wings resulting in positive $h_4$. However, seeing smears in the VPs at $\sim \pm 0\rlap{.}''7$, which show rotation in opposite directions and thus broaden the low-velocity region of the central VP leading to negative $h_4$. Higher spatial resolution would yield positive $h_4$ for the central VP, see subsection 4.5 below.

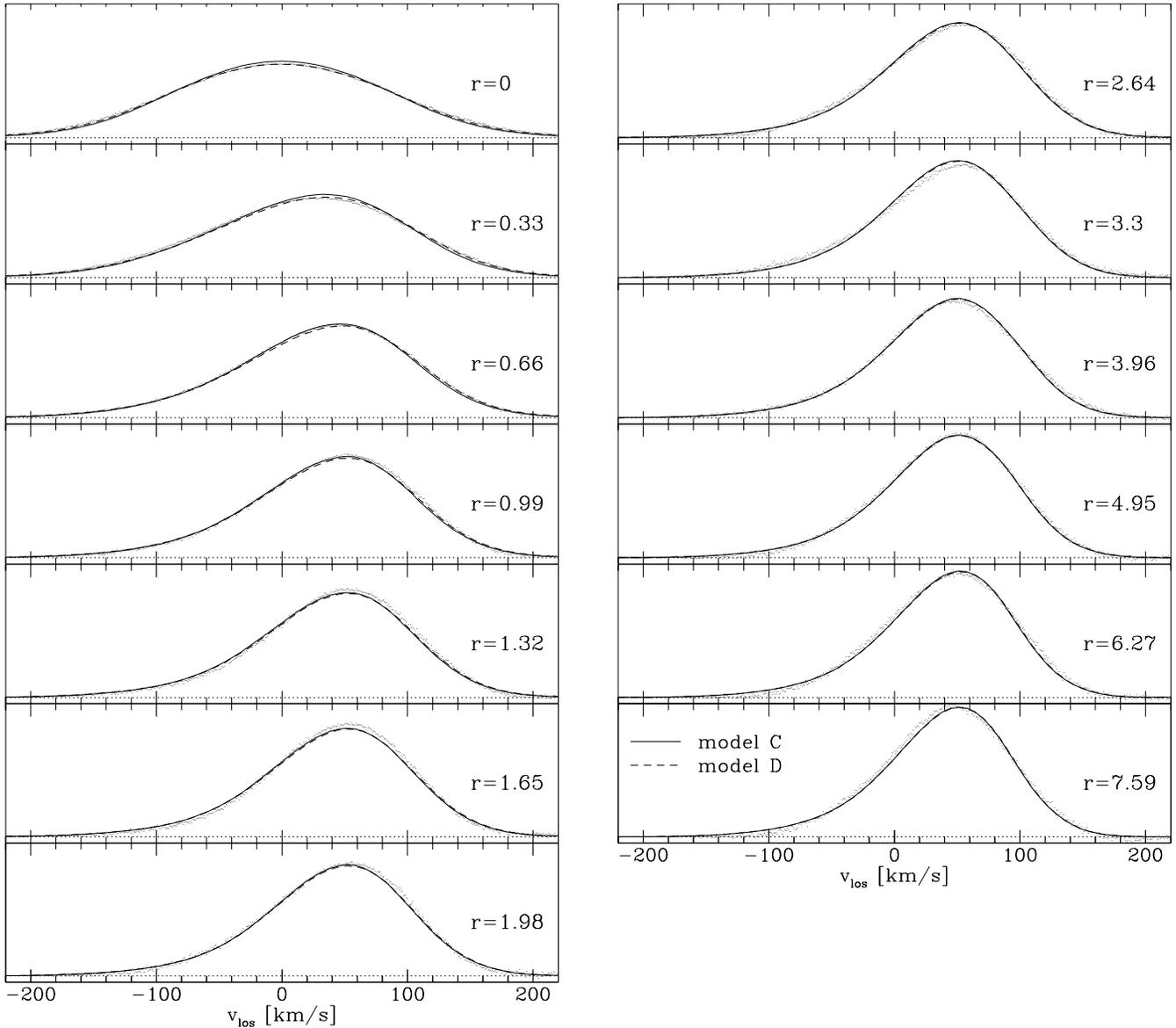

**Figure 7.** Comparing the VPs (normalized to unit area) along the major axis of the models with central black hole with those of M32 as observed by vdM94a. The latter are represented by points as in Fig. 5.

Second, the predicted $v_{\rm fit}$, $\sigma_{\rm fit}$ and $h_i$ differ slightly from the observed values, which is clearest on the major axis, where the errors are smallest. However, this seems to be hardly significant from the VPs in Fig. 7, where only the formal errors are considered and no systematic errors (e.g. template mismatching). Additionally, the VP-shape parameters $h_i$ are quite sensitive to $f_o$, as experiments with various $\alpha$ (equation 10) have shown, indicating that there might well be a $f_o$ which fits the VPs even better. It is interesting to note that the better fitting models have relatively less retrograde stars in the centre than outside about 7-10 pc$\approx$2-3$''$, i.e. rotate relatively more strongly in the region where the black hole reigns.

Third, there are few minor discrepancies, which may indicate that $f_o$ depends on $I_3$, in particular the rotation at slit position 4$''$ parallel to the major axis. The non-zero values of $h_3$ and $h_5$ on the minor axis, as well as the different rotation velocities at the same central distances on the two intermediate axes are even not consistent with any axisymmetric equilibrium model. However, it is not clear that these discrepancies are statistically significant.

The main conclusion is that a 2I-model with $i \approx 55°$ and $M_\bullet = 1.6\text{--}2 \times 10^6 M_\odot$ is mainly consistent with the kinematic data. This makes it unlikely, that any 3I-DF which fits the data equally well depends strongly on the third integral.

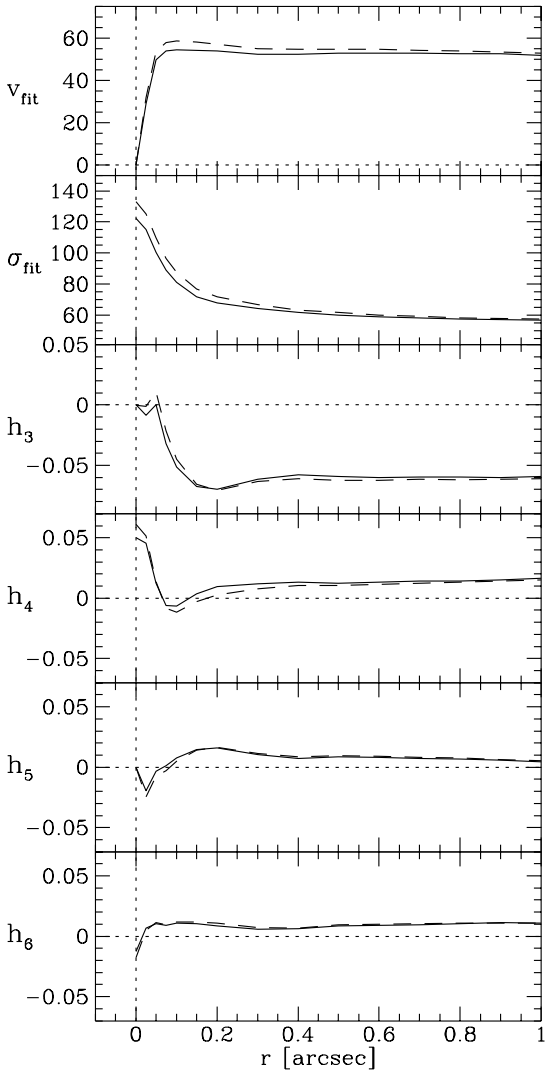

**Figure 8.** Predictions of the models with central black hole for observations with the $0''\!.09 \times 0''\!.09$ aperture of the Faint Object Spectrograph of the *HST*: the Gauss-Hermite fit parameters of the VPs along the innermost arcsecond of the major axis.

### 4.5 Predictions for observations with the *HST*

As has already been pointed out by Qian et al. (1994) *HST*-spectroscopy of the centre of M32 would be worthwhile, especially since the integration times necessary to gain high signal-to-noise ratios are short because of the high surface brightness. The models can be compared with such observations to further constrain the mass of a massive central object in M32 (note that the $0''\!.1$ resolution of the *HST* corresponds to $\approx 0.3$ pc at the distance of M32).

For models C and D (with black holes) I have computed the kinematics on the major axis inside $1''$, which would be measured using a square aperture of $0''\!.09 \times 0''\!.09$, the smallest available aperture on the *HST* Faint Object Spectrogaph. Figure 8 shows the corresponding values of the Gauss-Hermite-fit parameters along the major axis. At such a high resolution the central black hole would leave clear imprints on the kinematics: a high central velocity dispersion and strong gradient in projected rotation, as well as a central VP showing strong deviations from gaussian form – see also Fig. 9, where the predicted VPs at $0''$ and $0''\!.2$ are plotted. The high velocities in the neighbourhood of the black hole lead to broad wings of the central VP measured by values of $h_4 > 0.05$ (in contrast to the earth-based observations of vdM94a, for which $h_4 < 0$ in the centre, see the discussion above). A measurement of central VPs like those predicted here would further constrain the possible stellar dynamical configurations and provide more evidence for the presence of a black hole in M32. Another clue for a central dark mass would be the detection of stellar velocities which are well above the central escape velocity $v^\star_{esc}$ of the potential due to the stars alone. Such velocities would imply a deeper central potential and hence a dark component at least as compact as the stellar distribution – they cannot be explained by an extended dark halo, since then the associated low-binding-energy stars should also be visible at much larger radii. However, for M32 such a detection seems to be hardly possible, since the predicted fraction of stars with projected velocities exceeding $v^\star_{esc} \approx 387 \text{kms}^{-1}$ is very small, see Fig. 9.

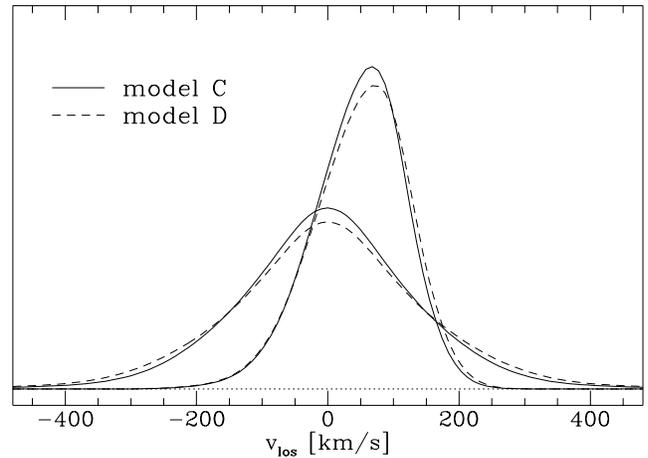

**Figure 9.** The major axis' VPs (normalized to unit area) at $0''$ and $0''\!.2$ that the models with central black hole predict for observations with the $0''\!.09 \times 0''\!.09$ aperture of the Faint Object Spectrograph aboard the *HST*.

### 4.6 Comparison with other recent work

Van der Marel et al. (1994b) used the mass model

$$\rho \propto m^{-1.435}(1 + [m/0''\!.55]^2)^{-0.423} \qquad (11)$$

with $m^2 = R^2 + (z/0.73)^2$ (assuming edge-on projection) to solve the Jeans equations up to third order, where a central black hole with mass $1.8 \times 10^6 M_\odot$ was added to the potential. Later on, Qian et al. (1994) have used the Hunter & Qian (1993) method to evaluate the 2I-DF for this model, and subsequently computed the VPs (which were not available from the pure Jeans equation model). This model differs from the successful models C and D above mainly in the four following points. It (i) assumes edge-on inclination, (ii) has a somewhat shallower cusp, (iii) has constant ellipticity, while M32 becomes rounder than E2.7 already at $10''$, and (iv) its surface brightness declines as $r^{-1.28}$ outside a few arcseconds, while that of M32 falls off steeper outside about 20″. Nevertheless, their results are very similar, even in some details, to those found in this work. However, there are two noticeable differences.

First, as for the edge-on model A above, they found too much motion on the minor axis and $VP_e$ on the major axis being less flat-topped than those observed. The authors interpret the latter as a suggestion that $f_e$ depends on $I_3$, while in the models presented here these discrepancies were removed by taking $i \approx 55°$. However Qian et al. claim this discrepancy in the $VP_e$ is hardly significant, in apparent conflict with Fig. 5.

The $h_i$-profiles of the major axis differ: $h_3$ of the Qian et al. model does not show a dip at 1″, as do the models presented here, and $h_4$ is smaller at all radii. Both differences may well be caused by the different forms for $f_o$ that have been used. However, in the centre there are very few retrograde stars in either model, and hence the freedom in $f_o$ is small. It seems possible that the discrepancy in $h_3$ is due to the shallower cusp of their model, since the steeper the density profile the more high-$L_z$ orbits a 2I-DF has to populate, and hence the more asymmetric the intrinsic and projected velocity distributions will be, leading to more negative $h_3$.

## 5 SUMMARY AND CONCLUSIONS

The Richardson (1972) –Lucy (1974) algorithm can be used to quickly and accurately compute the two-integral distribution function $f(E, L_z)$ for an axisymmetric stellar system; there are no restrictions regarding the functional form of the density $\rho(R, z)$ or otherwise, as long as density and potential can be represented on a grid, as is the DF itself. The computed 2I-DF is by construction positive and hence physical. Nevertheless, even unphysical DFs can be obtained as the difference of two positive ones. For better convergence of the iterative RL algorithm, it is very helpful, to transform the integral equation to be solved in an optimum form, in which the known projected quantity (e.g. the density) is as homogeneous as possible, and the kernel (e.g. the density of the orbits with same $E$ and $L_z$) is as local as possible (Appendix A).

In Section 3 it is shown, how this technique can be used to construct 2I-DFs for individual galaxies with the observed surface brightness distribution as the only input. Because of the uniqueness of $f_e$ and the simplicity of its evaluation (compared to 3I-DFs), the 2I-model can be used as a template model for the comparison of the kinematic data of the galaxy under investigation. In particular, it should be possible to tell whether the velocity distribution is less or more tangentially or radially anisotropic compared to that of the 2I-model, and if so, how strong the effects are, i.e. how strongly the DF depends on $I_3$.

In Section 4 the technique has been applied to the nearby elliptical galaxy M32. Starting from a composed surface brightness, which comprises the data of Kent (1987) and Peletier (1993) for $r > 4″$, the *HST* data of Lauer et al. (1992) for $0\rlap{.}″15 < r < 4″$ and a cusp $I \propto r^{-0.53}$ for $r < 0\rlap{.}″15$ as suggested by Lauer et al., the axisymmetric density was obtained for the inclination angles 55° and 90° (edge-on). For each of the resulting stellar distributions, I have first computed the 2I-DF without assuming a central black hole and have evaluated the projected kinematics (VPs), as if obtained under the observing conditions of vdM94a. The edge-on model was unable to fit the observed kinematics: it shows too little motion on the major axis in comparison to the minor axis, and significantly different shapes for the VPs on the major axis. The 55° inclined model (intrinsic E4.5) gave both correctly, but was unable to reproduce the high observed velocities inside 2″. To account for the latter, two additional models have been constructed with central point masses of $1.63 \times 10^6 M_\odot$ and $1.95 \times 10^6 M_\odot$. For the odd parts of the DFs a simple ansatz has been made. These models have kinematic observables, including the shapes of the VPs, in excellent agreement with those of M32 given by vdM94a. This means that the available kinematic data are consistent with a 2I-model with central black hole (or other compact dark mass) in the mass range from 1.6 to 2 million solar masses. It is not clear, whether three-integral models exist that equally well fit the observed kinematics, but contain a less or more massive black hole. Nevertheless, it seems unlikely, that such models would manage without any central dark point mass, for the following reason. To explain the high central dispersion and rotation without black hole, a highly radially anisotropic DF is required with the high central velocities occurring at the pericentres of stars on highly eccentric orbits with low binding energy. On the other hand, the kinematics of M32 outside of the innermost arcseconds indicate tangential anisotropy there (since a E4.5 2I-model fits the data), which is certainly not consistent with many stars being on highly eccentric orbits. From this argument I infer the presence of a dark mass of $1$–$3 \times 10^6 M_\odot$ in the centre of M32. Kinematic data at the resolution of the *HST* would give further constrains on the possible DFs of M32. For observations with the $0\rlap{.}″09 \times 0\rlap{.}″09$ aperture of the *HST* Faint Object Spectrograph the 2I-models with black hole predict a clearly non-gaussian central VP with broad wings, true and gauss-fit velocity dispersion of $151\,\mathrm{km\,s^{-1}}$ and $121\,\mathrm{km\,s^{-1}}$, respectively, for a black hole mass of $1.63 \times 10^6 M_\odot$, while a black hole of $1.95 \times 10^6 M_\odot$ would give $171\,\mathrm{km\,s^{-1}}$ and $131\,\mathrm{km\,s^{-1}}$.

The possibility to model not only the mean velocity and dispersion of the line-of-sight velocity distributions of galaxies, but also their shapes, the velocity profiles, is a great challenge. The additional information in the VPs constrains the possible dynamical configurations of the observed galaxies – as this paper demonstrates for M32 – and hence will lead to a better understanding of their internal structure. The main problem is the construction of dynamical models sophisticated enough to exploit the information in the VPs. The simplest such models are axisymmetric with $f = f(E, L_z)$. Only for very few galaxies have distribution functions been evaluated so far, and further work in this direction is certainly needed.

I anticipate making my FORTRAN codes for the evaluation of $f_e(E, L_z^2)$ and of its VPs available to the community within a few month after publication of this paper.

## ACKNOWLEDGMENTS

The author thanks Tim de Zeeuw and Roeland van der Marel for helpful discussions and submission of the RMS-velocities for M32. I am indebted to James Binney for many discussions and comments, which helped improving this paper. This research was financially supported by the PPARC.

## REFERENCES

Barnes J.E., Hernquist L., 1994, in preparation
Bender R., 1990, A&A 229, 441
Binney J.J., 1980, MNRAS 190, 873
Binney J.J., Tremaine S., 1987, Galactic Dynamics. Princeton Univ. Press, Princeton
Binney J.J., Davies R.L., Illingworth G.D., 1990, ApJ 361, 78
Dehnen W., 1993, MNRAS 265, 250
Dehnen W., Gerhard O.E., 1993, MNRAS 261, 311
Dehnen W., Gerhard O.E., 1994, MNRAS 268, 1019
Dressler A., Richstone D., 1988, ApJ 324, 701
Eddington A.S., 1916 MNRAS 76, 572
Evans N.W., 1994, MNRAS 267, 333
Franx M., Illingworth G.D., de Zeeuw P.T., 1991, ApJ 383, 112
Franx M., van Gorkom J.H., de Zeeuw P.T., 1994, ApJ, in press
Gerhard O.E., 1991, MNRAS 250, 812
Hunter C., Qian E., 1993, MNRAS 262, 401
Katz N., Gunn J.E., 1991, ApJ 377, 365
Kaufman L., 1987, IEEE Trans. Med. Imag., MI-6, 37
Kent S.M., 1987, AJ 94, 306
Kuijken K., Merrifield M.R., 1993, MNRAS 264, 712
Lauer T.R. et al., 1992, AJ 104, 552
Lucy L.B., 1974, AJ 79, 745
Lucy L.B., 1992, AJ 104, 1260
Lynden-Bell D., 1962, MNRAS 123, 447
Magorrian J., Binney J.J., 1994, MNRAS, submitted
Merritt D., Tremblay B., 1994, AJ 108, 514
Newton A.J., Binney J.J., 1987, MNRAS 210, 771
Palmer P.L., 1994, MNRAS 266, 697
Peletier R., 1993, A&A 271, 51
Qian E., de Zeeuw P.T., van der Marel R.P., Hunter C., 1994, MNRAS submitted
Quinlan G.D., Hernquist L., Sigurdsson S., 1994, ApJ submitted
Richardson W.H., 1972, J. Opt. Soc. Am. 62, 55
Richstone D., Bower G., Dressler A., 1990, ApJ 353, 118
Rix H.W., White S., 1992, MNRAS 254, 389
Rybicki G.B., 1987, in de Zeeuw P.T., ed, IAU Proc. 127, Structure and Dynamics of Elliptical Galaxies. Reidel, Dordrecht, p. 397
Shepp L.A., Vardi Y., 1982, IEEE Trans. Med. Imag., MI-1, 113
Tonry J.L., 1987, ApJ 322, 632
van der Marel R.P., 1991, MNRAS 253, 710
van der Marel R.P., Franx M., 1993, ApJ 407, 525
van der Marel R.P., Rix H.W., Carter D., Franx M., White S.D.M., de Zeeuw T.P., 1994a, MNRAS 268, 521 (vdM94a)
van der Marel R.P., Evans, N.W., Rix H.W., White S.D.M., de Zeeuw T.P., 1994b, MNRAS in press
Winsall M.L., Freeman K.C., 1993, A&A 268, 443
Young P., 1980, ApJ 242, 1232

## APPENDIX A: THE MODIFIED RICHARDSON-LUCY ALGORITHM

### A1  The original algorithm

Richardson (1972) and Lucy (1974) described an algorithm (hereafter RL algorithm) to solve the projection equation

$$\varphi(x) = \int \psi(\xi) K(x|\xi) \, d\xi \tag{A1}$$

for the unknown (intrinsic) $\psi(\xi)$ given the known projected (observed) function $\varphi(x)$ and the kernel $K(x|\xi)$, which all have to be non-negative and normalizable. Here the co-ordinates $x$ and $\xi$ may well be multi-dimensional. The scheme consists of successive applications of the two iterative steps:

$$\varphi_n(x) = \int \psi_n(\xi) K(x|\xi) \, d\xi \tag{A2}$$

$$\psi_{n+1}(\xi) = \psi_n(\xi) \frac{\int [\varphi(x)/\varphi_n(x)] K(x|\xi) \, dx}{\int K(x|\xi) \, dx}, \tag{A3}$$

where the subscript $n$ indexes the iteration step. In equation (A3) the integral in the observed domain includes all points $x$ to which $\psi(\xi)$ contributes. This algorithm converges in the sense that the logarithmic likelihood of the projected function,

$$H_n = \int \varphi(x) \log[\varphi_n(x)] \, dx, \tag{A4}$$

is maximized under the constraint $\int \varphi(x) \, dx = \int \varphi_n(x) \, dx$ (Shepp & Vardi 1982).

### A2  Modifying the RL algorithm

The RL algorithm may easily be modified by manipulating the projection equation (A1) such that its solution remains unchanged. The simplest way to do so is by multiplying it on both sides by some positive function $\alpha(x)$, which is only restricted by the constraint, that the product $\alpha(x)\varphi(x)$ must remain normalizable[3]. This manipulation affects only the correction step (equation A3), which becomes

$$\psi_{n+1}(\xi) = \psi_n(\xi) \frac{\int [\varphi(x)/\varphi_n(x)] \alpha(x) K(x|\xi) \, dx}{\int \alpha(x) K(x|\xi) \, dx}. \tag{A5}$$

The such modified algorithm converges in the sense that

$$H_n(\alpha) = \int \alpha(x) \varphi(x) \ln[\varphi_n(x)] \, dx \tag{A6}$$

is maximized under the constraint $\int \varphi(x) \, dx = \int \varphi_n(x) \, dx$.

Such a modification is sensible if the given projected function $\varphi(x)$ is significantly inhomogeneous, since in this case $H_n$ (equation A4) is dominated by points where $\varphi(x)$ is greatest, with the result that in regions where $\varphi(x)$ is small the original RL algorithm tends to converge rather slowly. If, for instance, $\varphi(x)$ is nowhere zero, a suitable choice is $\alpha(x) = \varphi(x)^{-1}$, which results in a scheme maximizing $\int dx \ln[\varphi_n(x)/\varphi(x)]$.

Another problem often encountered with the RL algorithm is the fitting of noise, which typically occurs after the first few iterations. Numerically, the unknown $\psi$ to be found is always represented on some grid, and hence only given with finite resolution. Therefore, one may smooth $\psi$ after each correction and before the projection step with a smoothing scale similar to the resolution element of the grid. This will guarantee a smooth result and avoid the fitting of small-scale noise.

---

[3] Another method would be to differentiate the projection equation with respect to some components of $x$ (which in the simplest case $\varphi = \int_0^x \psi \, d\xi$ already gives the solution) or any combination of this with multiplication. However, this method is useful for the RL algorithm only if the left hand side of the new projection equation is everywhere non-negative. In practice the accurate evaluation of the involved derivatives is likely to be problematic.

### A3  Accelerating the RL algorithm

For the purpose of speeding up the convergence, at each iteration the new guess for the intrinsic function $\psi_{n+1}(\xi)$ is written as $\psi_n(\xi) + \delta_\psi(\xi)$, where $\delta_\psi(\xi)$ is the correction of $\psi_n(\xi)$, which projects into $\delta_\varphi(x) = \int \delta_\psi(\xi) K(x|\xi) d\xi$. For accelerating the convergence of the algorithm, $\delta_\psi(\xi)$ is multiplied by a factor $\nu$, which after projection is chosen to be optimal and to obey $1 \leqslant \nu < f\nu_{\max}$, where $\nu_{\max}$ is given by the condition that $\psi_{n+1}(\xi)$ must not become zero and $f \leqslant 1$ is a fudge factor.

Usually, the optimal $\nu$ is determined by maximizing the resulting $H_{n+1}$ (Kaufman 1987; Lucy 1992). This can be done by numerically solving $dH_{n+1}/d\nu = 0$, which however, again involves an iterative procedure. Another possibility is, to choose the optimum $\nu$ by minimizing the relative RMS-error $\chi$ between $\varphi$ and $\varphi_{n+1}$, which yields

$$\nu_{\rm opt} = \sum \frac{(\varphi - \varphi_n)\delta_\varphi}{\varphi^2} \Bigg/ \sum \left(\frac{\delta_\varphi}{\varphi}\right)^2, \qquad (A7)$$

where the sums go over all data points.

Clearly, the function $\psi_H$ which maximizes the likelihood $H$ will be distinct from the function $\psi_\chi$ which minimizes the relative RMS-error, and therefore the choice of $\nu$ as in equation (A7) may retard or even stop the convergence. However, for the first iteration steps $\psi_n$ will differ clearly from both functions $\psi_H$ and $\psi_\chi$ when compared to their difference, and hence maximizing $H$ or minimizing $\chi$ will be quite similar. I found this choice of $\nu_{\rm opt}$ to work quite well, especially if the RL algorithm is suitably modified, which gives $\psi_H \approx \psi_\chi$.

## APPENDIX B: ALTERNATIVE RL SCHEMES TO RECOVER $f_{\rm e}(E, L_z^2)$

As mentioned in Section 2, one can apply the RL algorithm on the basic integral equation to be solved after taking a derivative on both of its sides, which reduces the dimensionality of the involved integrals by one, and hence should result in numerically faster RL schemes to recover $f_{\rm e}(E, L_z^2)$ than that given in Section 2. However, the accurate knowledge of the derivatives involved is required.

### B1  Scheme 1: taking the derivative $\partial/\partial \Psi$

Taking the partial derivative with respect to $\Psi$ on both sides of equation (1) results in the integral equation

$$A(\Psi, R) \equiv \frac{\partial \rho(\Psi, R)}{\partial \Psi} = 2^{3/2}\pi \int_0^\Psi dE \, \frac{f_{\rm e}(E, 2R^2[\Psi - E])}{\sqrt{\Psi - E}}. \quad (B.1)$$

The modified RL algorithm gives for the correction step

$$\frac{f_{e,n+1}(E, L_z^2)}{f_{e,n}(E, L_z^2)} = \frac{\int_{\Psi_1}^{\Psi_2} \frac{d\Psi}{\sqrt{\Psi - E}} \left[\frac{A(\Psi, R)^{1-p}}{A_n(\Psi, R)}\right]_{R = L_z/\sqrt{2(\Psi-E)}}}{\int_{\Psi_1}^{\Psi_2} \frac{d\Psi}{\sqrt{\Psi - E}} A(\Psi, \frac{L_z}{\sqrt{2(\Psi-E)}})^{-p}} \quad (B.2)$$

for $L_z \neq 0$, and

$$\frac{f_{e,n+1}(E, 0)}{f_{e,n}(E, 0)} = \frac{\int_E^{\Psi(0,0)} \frac{d\Psi}{\sqrt{\Psi - E}} A(\Psi, 0)^{1-p}/A_n(\Psi, 0)}{\int_E^{\Psi(0,0)} \frac{d\Psi}{\sqrt{\Psi - E}} A(\Psi, 0)^{-p}} \quad (B.3)$$

for $L_z = 0$. $\Psi_1$, $\Psi_2$ are given in Section 2, while the exponent $p$ can be chosen to optimize the convergence.

### B2  Scheme 2: taking the derivative $\partial/\partial R$

Multiplying equation (1) with $R$ and subsequently taking the partial derivative with respect to $R$ on both sides yields

$$B(\Psi, R) \equiv \frac{\partial (R\rho[\Psi, R])}{\partial R}$$
$$= 2^{5/2}\pi \int_0^\Psi dE \, \sqrt{\Psi - E} \, f_{\rm e}(E, 2R^2[\Psi - E]). \quad (B.4)$$

The correction step of the modified RL algorithm reads

$$\frac{f_{e,n+1}(E, L_z^2)}{f_{e,n}(E, L_z^2)} = \frac{\int_{\Psi_1}^{\Psi_2} d\Psi \left[\frac{B(\Psi, R)^{1-p}}{B_n(\Psi, R)}\right]_{R = L_z/\sqrt{2(\Psi-E)}}}{\int_{\Psi_1}^{\Psi_2} d\Psi \, B(\Psi, \frac{L_z}{\sqrt{2(\Psi-E)}})^{-p}} \quad (B.5)$$

for $L_z \neq 0$ and is given by equation (3) for $L_z = 0$.

## APPENDIX C: COMPUTING VELOCITY PROFILES

The velocity profile (VP) is given by a three-dimensional integral over the distribution function, namely over the two velocity components in the plane of sky and along the line of sight. However, if one wants to compare with observed VPs, effects of binning and seeing have to be included, each of them resulting in another two-dimensional integration. The best way of handling such high dimensional integrals is a Monte Carlo integration, a straightforward application of which gives a highly ineffective procedure for the evaluation of a VP, since most of the randomly chosen phase space points lie outside the region which dominates the VP, i.e. at small or even negative binding energies.

To get a more efficient procedure one needs to preferentially pick out phase space points resulting in large values of the DF. In Monte Carlo integration technique this is called "reduction of variance", it is equivalent to a substitution of a variable of integration, i.e. one has to multiply with the corresponding Jacobian. After some experiments I ended up with the iteration of the following steps:

(1) Set $F = 1$, the factor containing the Jacobians due to the reductions of variance and the DF.

(2) Choose a position on the sky $(x, y)$ at random out of the binning area, and add a two dimensional vector drawn randomly from the point spread function. Both times one may well prefer small radii to account for the gradient in surface brightness, in which case $F$ has to be multiplied by the corresponding reduction factors.

(3) Choose a position $s$ on the line of sight at random out of $0 \leqslant s/(s + \alpha) < 1$; $\alpha = \sqrt{x^2 + y^2 + a^2}$, where $a$ is a small number, and $s = 0$ corresponds to the point, where a pure spheroidal density distribution with similar ellipticity would be maximal along the line of sight; compute the galaxy's intrinsic co-ordinates $(R, z)$. To correct for the preference of small $s$, the factor $F$ has to be multiplied by $(s+\alpha)^2/\alpha$.

(4) Choose energy randomly out of $0 < E^2 \leqslant \Psi(R,z)^2$, calculate $v = \sqrt{2(\Psi - E)}$, and multiply $F$ by $\Psi(R,z)^2/(2E)$.

(5) Choose $v_\phi$ randomly out of $-v \leqslant v_\phi \leqslant v$, calculate $v_m = \sqrt{v^2 - v_\phi^2}$, evaluate $f(E, Rv_\phi)$, and multiply $F$ with $v f(E, Rv_\phi)$

(6) Loop over $0 \leqslant \psi < 2\pi$, for each $\psi$ set $v_R = v_m \cos\psi$, $v_z = v_m \sin\psi$, compute $v_{\mathrm{los}}$, and add $F$ to the corresponding pixel of the VP-histogram.

Points (4) to (6) are based on the identity

$$\int_{v^2 \leqslant 2\Psi} \mathrm{d}^3 v = \int_0^{\Psi^2} \frac{\mathrm{d}E^2}{2E} \int_{-\sqrt{2(\Psi-E)}}^{\sqrt{2(\Psi-E)}} \mathrm{d}v_\phi \int_0^{2\pi} \mathrm{d}\psi. \qquad (\mathrm{C}.1)$$

The most time consuming piece in this whole procedure is the evaluation of the DF, since it involves a two-dimensional interpolation. The trick in the above algorithm is the loop in step (6), which exploits the fact, that a 2I-DF is independent on the angle $\psi$ in velocity space. This allows the number of evaluations of $f(E, L_z)$ to be much smaller than the number of points actually added to the histogram.